\documentclass[twocolumn,aps,showpacs,preprintnumbers,amsmath,amssymb,byrevtex,superscriptaddress]{revtex4}
\usepackage{graphicx}% Include figure files
\usepackage{dcolumn}% Align table columns on decimal point
\usepackage{bm}% bold math

\begin{document}
\preprint{Fluorescence Spectrum}
\title{Fluorescence Spectra of a Two-Level Atom Embedded in a Three-Dimensional Photonic Crystal}
\author{Ray-Kuang Lee}
\affiliation{Institute of Electro-Optical Engineering, National Chiao-Tung University, Hsinchu, Taiwan}
\affiliation{National Center for High-Performance Computing, Hsinchu, Taiwan}
\author{Yinchieh Lai}
\email{yclai@mail.nctu.edu.tw}
\affiliation{Institute of Electro-Optical Engineering, National Chiao-Tung University, Hsinchu, Taiwan}
\date{\today}

\begin{abstract}
Steady-state fluorescence spectra of a two-level atom embedded in a three-dimensional photonic bandgap crystal and driven by a monochromatic classical electrical field is calculated theoretically for the first time as we know.
The non-Markovian noises caused by the non-uniform distribution of photon density of states near the photonic bandgap are handled by a new approach in which the Liouville operator expansion is utilized to linearize the generalized optical Bloch equations.
The fluorescence spectra are then directly solved by the linearized Bloch equations in the frequency domain.
We find that if the atomic energy level is far from the bandgap, fluorescence spectra with Mollow's triplets are observed.
However, when the atomic energy level is near the bandgap, the relative magnitude and the number of the fluorescence peaks are found to be varied according to the wavelength offset.

\end{abstract}

\pacs{32.50.+d, 42.70.Qs, 42.50.Pq}

\keywords{Fluorescence-atoms, Photonic bandgap materials, Cavity quantum electrodynamics}

\maketitle

\section{Introduction}
In recent years, with the advance of new fabrication technologies, it has become more feasible to actually utilize higher dimensional periodic dielectric structures (or especially the photonic bandgap crystals) \cite{SJohn87,EYablonovitch87} for modifying the properties of the photon states as well as the properties of the spontaneous emission.
Like electrons in solid state crystals, photons are prohibited to propagate inside photonic bandgap crystals due to the lack of available photon states.
In general, the photon density of states (DOS) of photonic crystals has a highly non-uniform distribution near the bandgap, which is totally different from the case in free space.
Such a non-uniform distribution of the DOS has been investigated by many authors and has provided a new and experimentally feasible platform for investigating photon-atom interaction.
Many new phenomena such as photon-atom bound states \cite{SJohn90}, spectral splitting \cite{SJohn94}, quantum interference dark line effect \cite{SZhu97}, phase control of spontaneous emission \cite{EPaspalakis98}, transparency near band edge \cite{EPaspalakis99}, and single-atom switching \cite{MFlorescu01} have been theoretically discovered in the presence of the bandgap.
From the Aulter-Townes spectra for atoms coupled to a photonic bandgap structure \cite{SJohn94} or equivalently a frequency-depended photon DOS \cite{MLewenstein88}, the modification of the spontaneous emission caused by the environment \cite{EPurcell46} actually can be verified. However, all of the above studies only focused on the transient behavior of the atom-photon interactions and to the best of our knowledge there is still no theoretical treatment on calculating the steady-state fluorescence spectra in photonic bandgap crystals.

To investigate this problem, the approach of the present paper is to treat the photon states of the photonic crystal as the background reservoir and introduce non-Markovian noise operators caused by the non-uniform DOS distribution near the band edge.
To model the non-uniform DOS distribution near the bandgap, an isotropic dispersion relation \cite{SJohn94} and an anisotropic dispersion relation \cite{SZhu00} have been proposed to serve as a simple but qualitatively correct model for theoretical analyses.
With the use of the three-dimensional anisotropic dispersion relation, we will investigate the steady-state properties of the resonance fluorescence spectra emitted by a two-level atom which is embedded in a photonic bandgap crystal and driven by a classical pumping light. 
Due to the non-Markovian nature of the atom-field interaction in this case, we can not directly utilize the Born-Markovian approximation which is usually used in quantum optics for treating atom-photon interaction problems.
To overcome this difficulty, we first derive the generalized optical Bloch equations without any approximation and then use the zero-order Liouville operator expansion to approximate the nonlinear terms. 
By solving the simplified linear equations in the Fourier domain directly we can calculate the stationary two-time correlation functions of the atomic operators as well as the spectral distribution of the resonance fluorescence. 
When the emission frequency of the atom is far from the bandedge, the triplet spectral shape is obtained, just as one will expect intuitively from the results first calculated by B. Mollow \cite{BMollow69} for the free space case. 
But when we change the emission wavelength of the atom to be close to the band edge, both the number of the peaks as well as their profiles are found to be varied depending on the wavelength offset between the atomic transition wavelength and the band edge. The details of these results will be reported in the rest of this paper.

The paper is organized as follows: in section \ref{secBloch} we derive the generalized optical Bloch equations with noise operators caused by the surrounding reservoir.
In section \ref{secPBG} we use the anisotropic dispersion relation for modeling the photon DOS of the three-dimensional photonic bandgap structure and based on this model, we present the calculated fluorescence spectra in section \ref{secFRS}. Finally, a brief conclusion is given in section \ref{secConclu} and the first order Liouville operator expansion is carried out in the appendix for accuracy checking.

\section{Theoretical Model}
\label{secBloch}

With the help of rotation wave approximation, one can use the Jaynes-Cummings model to describe the atom-photon interaction \cite{MQO}.
For a single two-level atom embedded in a photonic bandgap crystal and driven by a monochromatic classical pump light, the photon states in the photonic bandgap crystal can be treated as the background reservoir.
The Hamiltonian of the system can be written as: 
\begin{eqnarray} 
\label{Hamiltonian} 
H &=& \frac{\hbar}{2}\omega_a\sigma_z+\hbar\sum_k\omega_k a_k^\dag a_k +\frac{\Omega}{2}\hbar(\sigma_-e^{i\omega_L t}+\sigma_+e^{-i\omega_L t})\nonumber\\
&+&\hbar\sum_k(g_k \sigma_+ a_k + g_k^\ast a_k^\dag \sigma_-) 
\end{eqnarray} 
where the transition frequency of the atom and the frequency of the pumping light are denoted by $\omega_a$ and $\omega_L$ respectively, $a_k^\dag$ and $a_k$ are the creation and annihilation operators of the background field reservoir, $\Omega $ is the Rabi-flopping frequency of the atom under the external pumping light and also represents the relative magnitude of the pumping light, $\sigma_z\equiv(|2\rangle\langle 2|-|1\rangle\langle 1|)$, $\sigma_+\equiv|2\rangle\langle 1|=\sigma_-^\dag$ are the usual Pauli matrices for a two-level atom, and $g_k$ is the atom-field coupling constant.
Here we use the index $k$ to label different photon states.
The form of the coupling constant $g_k$ depends on the gauge one choose.
Although the Lamb shifts calculated in non-relativistic quantum field theories are different with the $\hat{p}\cdot \hat{A}$ and $\hat{r}\cdot\hat{E}$ formula \cite{WLamb87}, one can use the $\hat{p}\cdot\hat{A}$ formulation to get the correct form for the atom-field coupling \cite{GKweon95}, if the electromagnetic field varies little over the spatial extent of the electronic wave function.
Under the $\hat{p}\cdot\hat{A}$ formulation the coupling constant $g_k$ can be expressed as: 
\begin{eqnarray} 
\label{eqgk}
g_k(\hat{\textbf d},\vec{r}_0)\equiv g_k= |d| \omega_a \sqrt{\frac{1}{2\hbar \epsilon_0\omega_k V}}\,\hat{\textbf d}\cdot {\textbf E}^{\ast}_k (\vec{r}_{0}) 
\end{eqnarray} 
Here we use notations $|d|$ for the magnitude of the atomic dipole moment, $\hat{\textbf d}$ for the unit vector along the direction of the dipole moment, $V$ for the volume of the quantization space, and $\epsilon_0$ for the Coulomb constant.

The evolution equations for the atomic operators can be derived directly from the Hamiltonian in Eq. (\ref{Hamiltonian}).
By transferring the system to the rotating frame with the frequency $\omega_a$ and eliminating the reservoir field operators, we get the generalized Bloch equations as follows: 
\begin{eqnarray} 
\label{es1}
\dot{\sigma}_-(t) &=& i\frac{\Omega}{2}\sigma_z(t)e^{-i\Delta t}\\\nonumber
&+& \int_{-\infty}^t d\,t' G(t-t')\sigma_z(t)\sigma_-(t') + n_-(t)\\ 
\label{es2} 
\dot{\sigma}_+(t) &=& - i\frac{\Omega}{2}\sigma_z(t)e^{i\Delta t}\\\nonumber
&+& \int_{-\infty}^t d\,t' G_c(t-t')\sigma_+(t')\sigma_z(t) + n_+(t)\\ 
\label{es3} 
\dot{\sigma}_z(t) &=& i\Omega(\sigma_-(t)e^{i\Delta t} -\sigma_+(t)e^{-i\Delta t} + n_z(t)\\\nonumber
&&\hspace{-1.5cm} -2\int_{-\infty}^tdt' [G(t-t')\sigma_+(t)\sigma_-(t') + G_c(t-t')\sigma_+(t')\sigma_-(t)]
\end{eqnarray} 
Here $\Delta\equiv\omega_L-\omega_a$, and $\Delta_k\equiv\omega_a-\omega_k$.
The two functions, $G(\tau)$ and $G_c(\tau)$, are the memory functions due to atom-reservoir interaction and are defined as  $G(\tau) \equiv \sum_k |g_k|^2 e^{i\Delta_k\tau}\Theta(\tau)$, and $G_c(\tau) \equiv \sum_k |g_k|^2 e^{-i\Delta_k\tau}\Theta(\tau)$. Here $\Theta(\tau)$ is the Heaviside step function. 
Moreover, the three noise operators $n_-(t)$, $n_+(t)$, and $n_z(t)$ are expressed as follows: 
\begin{eqnarray} 
&&n_-(t) = i\sum_k g_k e^{i\Delta_k t}\sigma_z(t) a_k({-\infty})\\
&&n_+(t) = - i\sum_k g_k^\ast e^{-i\Delta_k t}a_k^\dag({-\infty}) \sigma_z(t)\\
&&n_z(t) = -2 i\sum_k g_k e^{i\Delta_k t}\sigma_+(t) a_k({-\infty})\\\nonumber
&&\hspace{1cm} + 2 i\sum_k g_k^\ast e^{-i\Delta_k t}a_k^\dag({-\infty}) \sigma_-(t) 
\end{eqnarray} 
Supposing that the reservoir is in thermal equilibrium, then the mean and correlation of the reservoir field operators before interaction will be: 
\begin{eqnarray}
&&\langle a_k({-\infty})\rangle_R = \langle a_k^\dag({-\infty})\rangle_R = 0\\
&&\langle a_k({-\infty}) a_{k'}({-\infty})\rangle_R = 0\\
&&\langle a_k^\dag({-\infty}) a_{k'}^\dag({-\infty})\rangle_R = 0\\
&& \langle a_k^\dag({-\infty}) a_{k'}({-\infty})\rangle_R = \bar{n}_k \delta_{kk'}\\
&& \langle a_k({-\infty}) a_{k'}^\dag({-\infty})\rangle_R = (\bar{n}_k+1) \delta_{kk'} 
\end{eqnarray} 
Here $\bar{n}_k$ is the mean quantum numbers of the reservoir modes under thermal equilibrium.

Using the statistical characteristics of the reservoir field operators, it can be easily shown that the three noise operators $n_-(t)$, $n_+(t)$, and $n_z(t)$ are zero mean with their correlation functions given below:
\begin{eqnarray*} 
&&\hspace{-0.4cm}\langle n_-(t)\rangle_R = \langle n_+(t)\rangle_R = \langle n_z(t)\rangle_R =0 \\ 
&&\hspace{-0.4cm}\langle n_-(t)n_-(t')\rangle_R = \langle n_+(t)n_+(t')\rangle_R = 0\\
&&\hspace{-0.4cm}\langle n_-(t)n_+(t')\rangle_R = \sum_k |g_k|^2 (\bar{n}_k+1) e^{i\Delta_k(t-t')} \langle\sigma_z(t)\sigma_z(t')\rangle \\ 
&&\hspace{-0.4cm}\langle n_+(t)n_-(t')\rangle_R = \sum_k |g_k|^2 \bar{n}_k e^{-i\Delta_k(t-t')} \langle\sigma_z(t)\sigma_z(t')\rangle \\ 
&&\hspace{-0.4cm}\langle n_z(t)n_z(t')\rangle_R = 4 \sum_k |g_k|^2 [(\bar{n}_k+1) e^{i\Delta_k(t-t')} \langle\sigma_+(t)\sigma_-(t')\rangle\\\nonumber
&&\hspace{1.8cm}+\, \bar{n}_k e^{-i\Delta_k(t-t')} \langle\sigma_-(t)\sigma_+(t')\rangle] 
\end{eqnarray*} 
Since in general the correlation functions of these noise operators are not delta correlated at time (non-Markovian), we cannot directly use the Born-Markovian approximation to solve the problem.
One can see that the correlation functions depend not only on the photon density of states, but also on the correlations of the atomic operators.
Eqs. (\ref{es1}-\ref{es3}) are called the generalized optical Bloch equations and will serve as the starting point for our further derivation. 

\section{Modeling of Three-Dimensional Photonic Crystals} 
\label{secPBG}
Due to the bandgap characteristics of photonic crystals, the spectral distribution of the photon DOS will have one or more discontinuities and is highly non-uniform near the bandedge.
Although in general the DOS of photonic bandgap crystals will vary with the geometrical structure and the dielectric constants of the materials they are made by, it is still possible to introduce some simple formula for approximately describing the DOS behavior near the bandgap.
In the literature an isotropic dispersion relation \cite{SJohn94} and an anisotropic dispersion relation \cite{SZhu00} have been proposed to serve as a simple but qualitatively correct model for theoretical analyses.
However, according to the results from a full vectorial numerical calculation \cite{ZLi00}, the DOS near the bandgap for a three-dimensional photonic crystal increases from zero and behaves more like the anisotropic model. This is why we will adopt the anisotropic model in the following derivation.
From Eq. (\ref{eqgk}), one knows that the magnitude of the coupling constant is also dependent on the local electric field.
Although in general one should apply this position-dependent coupling constant with the actual DOS to fully describe the photon-atom interaction within a photonic bandgap crystals, however for simplicity we will use the anisotropic model and a constant coupling coefficient for all the field modes to perform the numerical calculation.

For three dimensional photonic bandgap crystals, if the wavevector that corresponds to the bandedge is ${\textbf k}_0^i$, then the dispersion relation in the anisotropic model is described by the following form:
$\omega_k=\omega_c+A|{\textbf k}-{\textbf k}_0^i|^2$, where $A$ is a model dependent constant and $\omega_c$ is the band edge frequency.
Based on this dispersion relation, the corresponding DOS is given by: $ D(\omega) = \frac{1}{A^{3/2}}\sqrt{\omega-\omega_c}\Theta(\omega-\omega_c)$.
The memory functions under the anisotropic model also can be derived as: 
\begin{eqnarray} 
\label{Gpbg1} 
\tilde{G}(\omega) &=& \beta^{3/2}\frac{-i}{\sqrt{\omega_c}+\sqrt{\omega_c-\omega_a-\omega}}\\ 
\label{Gpbg2} 
\tilde{G_c}(\omega) &=& \beta^{3/2}\frac{i}{\sqrt{\omega_c}+\sqrt{\omega_c-\omega_a+\omega}} 
\end{eqnarray} 
where $\beta^{3/2}=\frac{\omega_a^2 d^2}{6\hbar\epsilon_0\pi A^{3/2}}\eta$, and we have used the space average coupling strength $\eta\equiv\frac{3}{8\pi}\int d\Omega|\hat{\textbf d}\cdot{\textbf E}|^2$ in the derivation.

\begin{figure} 
\begin{center}
\includegraphics[width=3.0in]{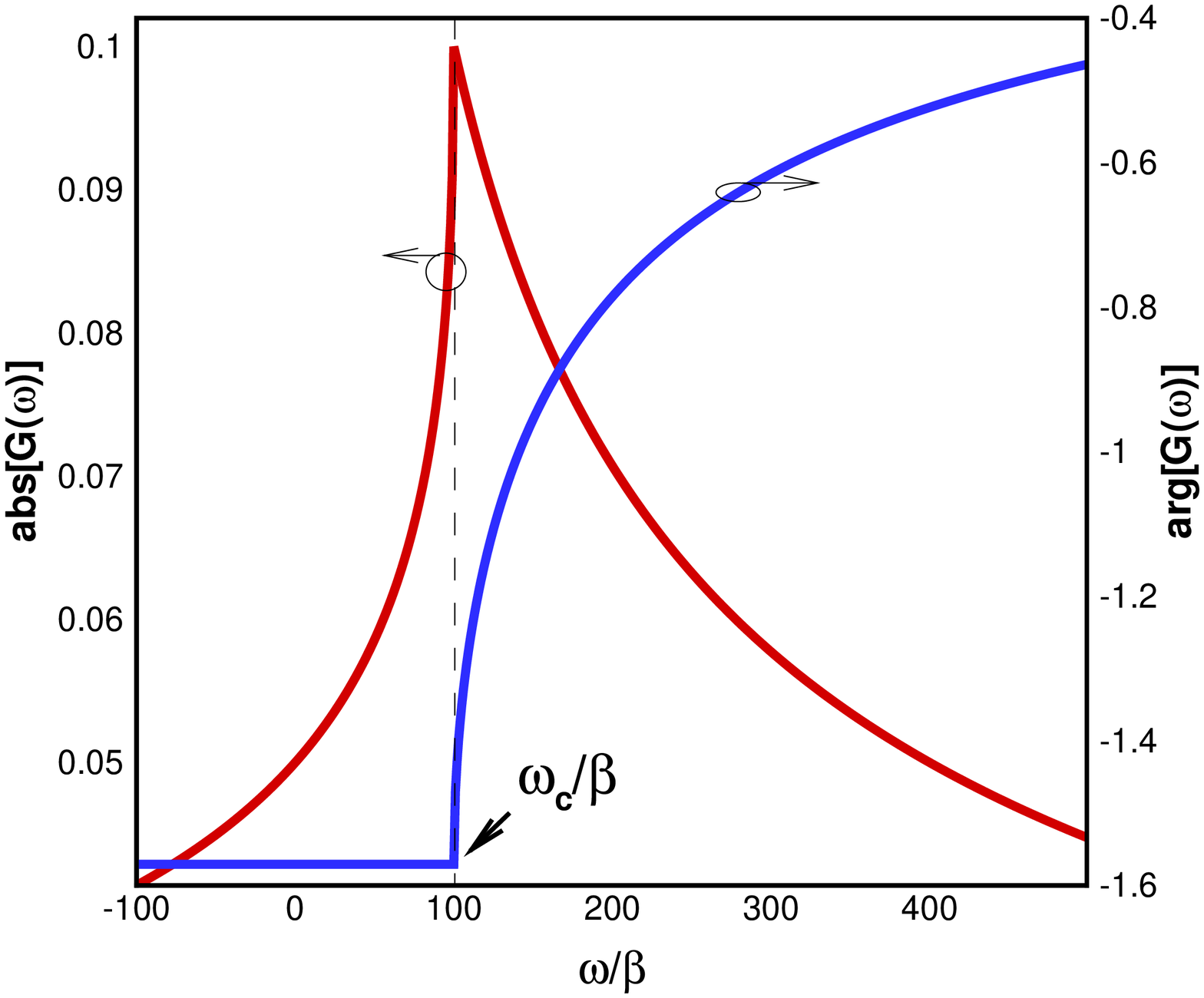} 
\caption{Amplitude and phase spectra of the memory function $G(\omega)$ with $\omega_c = 100 \beta$. The memory function is non-uniform around the bandedge and becomes pure imaginary inside the bandgap.} 
\label{figgw}
\end{center}
\end{figure}

From Fig. \ref{figgw}, we can see that the spectrum of the memory function $G(\omega)$ are non-uniform and asymmetric as we expect.
When the frequency is below the bandedge frequency $\omega_c$, the memory function becomes pure imaginary, indicating the inhibition of the spontaneous emission inside the bandgap.
The spectrum for another memory function $G_c(\omega)$ is also similar.
It can be easily checked that the full-width-half-maximum (FWHM) bandwidth of the memory functions in Eq. (\ref{Gpbg1}) and  Eq. (\ref{Gpbg2}) are $4 \omega_c$. 
For the bandgap in optical domain, the order of $\omega_c$ is about $10^{14-15}$ Hz, and the typical lifetime of the atom is from $10^{-3}$ sec to $10^{-9}$ sec, which is much longer than the response time of the memory functions.
Therefore it should be possible to approximate the two-time operator products in Eqs. (\ref{es1}-\ref{es3}) by the equal time operator products with the introduction of the Liouville operator expansion to be given below.

For a two-level atom system described by the Hamiltonian $H$, the time evolution of the atomic operators can be written in general as:   
\begin{eqnarray*}
\sigma_{ij}(t)\equiv e^{-i{\cal L}(t-t')}\sigma_{ij}(t')=\sum_{n=0}^\infty \frac{[-i(t-t')]^n}{n!}{\cal L}^n\sigma_{ij}(t') 
\end{eqnarray*} 
where the Liouville operator ${\cal L}$ is defined as 
\begin{eqnarray} 
\label{Liouville} 
{\cal L}^n \sigma_{ij}(t')=\frac{1}{\hbar^n}[\cdots[\sigma_{ij}(t'),H],H],\cdots,H] 
\end{eqnarray} 
In this paper, we will only consider the case in which the atom is with a longer lifetime and the pumping is not extremely high (i.e., the inverse of the Rabi frequency is much larger than the time scale of the memory functions). In this way we can be sure that the time scale of the atomic evolution will be always much longer than the time scale of the memory functions.
Under such assumptions, it should be legitimate for us to simply apply the zero-th order perturbation terms. This is equivalent to use the equal time operator products to replace the two-time operator products.
We have also checked the accuracy of the results by including the first order perturbation terms. The formulation is given in the Appendix and the numerical results only show difference when the intensity pumping is extremely high. 

It should be noted that the approximation we have made is valid only when the time scale of the system response is much slower compared to the time scale of the memory function. However, with the use of our approximation, we still keep the finite response time of the memory function in the formulation. This allows us to study the effects that are not considered in the usual Markovian regime, where the memory function is simply approximated by a delta-function. Although our approximation includes only one portion of the non-Markovian nature of the problem, it should still be quite valid for the fluorescence spectrum calculation considered in the present work since here the memory function time scale is typically much shorter than the atomic response time scale.

Under the zero-th order Liouville operator expansion approximation, $\sigma_z(t) \approx \sigma_z(t')$, and $\sigma_\pm(t) \approx \sigma_\pm(t')$, and with the identities of Pauli matrices, the generalized optical Bloch equations in Eqs. (\ref{es1}-\ref{es3}) can be reduced to: 
\begin{eqnarray*} 
&&\hspace{-0.4cm}\dot{\sigma}_-(t) = i\frac{\Omega}{2}\sigma_z(t)e^{-i\Delta t} -\int_{-\infty}^t d t' G(t-t')\sigma_-(t') + n_-(t)\\ 
&&\hspace{-0.4cm}\dot{\sigma}_+(t) = -i\frac{\Omega}{2}\sigma_z(t)e^{i\Delta t} -\int_{-\infty}^t d t' G_c(t-t')\sigma_+(t')+ n_+(t)\\ 
&&\hspace{-0.4cm}\dot{\sigma}_z(t) = i\Omega (\sigma_-(t)e^{i\Delta t} - \sigma_+(t)e^{-i\Delta t})\\\nonumber
&&\hspace{0.4cm}- \int_{-\infty}^t d t' [G(t-t')+G_c(t-t')](1+\sigma_z(t')) + n_z(t) 
\end{eqnarray*} 
Please note that these equations are now in a linear form  with the non-Markovian memory functions. 
By using Fourier transform, we can directly solve these modified optical Bloch equations as follows: 
\begin{eqnarray} 
\label{ssmatrix} 
\overline{\overline{\cal M}}(\omega)\cdot\vec{\cal X}(\omega) =\vec{\cal X}_0(\omega)
\end{eqnarray} 
where 
\begin{widetext}
\begin{eqnarray*}
&& \overline{\overline{\cal M}}(\omega) = \left(\begin{array}{ccc} -i(\omega+\Delta)+\tilde{G}(\omega) & 0 & - i\frac{\Omega}{2} \\0 & -i(\omega-\Delta)+\tilde{G_c}(\omega) & i\frac{\Omega}{2} \\ -i\Omega & i\Omega & - i\omega+\tilde{G}(\omega)+\tilde{G_c}(\omega) \end{array}\right)\\ 
&& \vec{\cal X}(\omega) = \left(\begin{array}{c} \tilde{\sigma}_-(\omega+\Delta) \\ \tilde{\sigma}_+(\omega-\Delta) \\ \tilde{\sigma_z}(\omega)\end{array}\right)\mbox{,}\quad \mbox{and} \quad\vec{\cal X}_0(\omega) = \left(\begin{array}{c} \tilde{n}_-(\omega+\Delta) \\ \tilde{n}_+(\omega-\Delta) \\ -2\pi[\tilde{G}(\omega)+\tilde{G_c}(\omega)]\delta(\omega)+\tilde{n}_z(\omega) \end{array}\right) 
\end{eqnarray*} 
\end{widetext}
and $\tilde{n}_-(\omega)$, $\tilde{n}_+(\omega)$, $\tilde{n}_z(\omega)$, $\tilde{G}(\omega)$, and $\tilde{G_c}(\omega)$ are Fourier transforms of $n_-(t)$, $n_+(t)$, $n_z(t)$, $G(t)$, and $G_c(t)$, respectively. The solutions of Eq. (\ref{ssmatrix}) are  
\begin{widetext}
\begin{eqnarray}
\label{ss1}
\tilde{\sigma}_-(\omega+\Delta) &=& \frac{(2 g_\omega h_\omega+\Omega^2)\tilde{n'}_{-\omega}+\Omega^2\tilde{n'}_{+\omega}+i\,g_\omega \Omega\tilde{n}_z(\omega)-i 2\pi g_\omega \Omega \tilde{G'}_\omega\delta(\omega)}{\Omega^2 (f_\omega+g_\omega)+2 f_\omega g_\omega h_\omega}\\
\label{ss2}
\tilde{\sigma}_+(\omega-\Delta) &=& \frac{\Omega^2 \tilde{n'}_{-\omega}+(2 f_\omega h_\omega+\Omega^2)\tilde{n'}_{+\omega} -i\,f_\omega\Omega\tilde{n}_z(\omega)+i 2\pi f_\omega \Omega \tilde{G'}_\omega\delta(\omega)}{\Omega^2 (f_\omega+g_\omega)+2 f_\omega g_\omega h_\omega}\\
\label{ss3}
\tilde{\sigma}_z(\omega) &=& \frac{2 i\,g_\omega \Omega\tilde{n'}_{-\omega}-2 i\,f_\omega\Omega\tilde{n'}_{+\omega}+2 f_\omega g_\omega\tilde{n}_z(\omega)-4\pi f_\omega g_\omega \tilde{G'}_\omega\delta(\omega)}{\Omega^2 (f_\omega+g_\omega)+2 f_\omega g_\omega h_\omega}
\end{eqnarray}
\end{widetext}

Here we have used the following shorthand notations:
\begin{eqnarray*}
&& f_\omega = f(\omega) \equiv -i\omega-i\Delta+\tilde{G}(\omega)\\
&& g_\omega = g(\omega) \equiv -i\omega+i\Delta+\tilde{G_c}(\omega)\\
&& h_\omega = h(\omega) \equiv - i\omega+\tilde{G}(\omega)+\tilde{G_c}(\omega)\\
&& \tilde{n'}_{\pm\omega} = \tilde{n'}_\pm (\omega) \equiv \tilde{n}_\pm(\omega\mp\Delta)\\
&& \tilde{G'}_\omega = \tilde{G'}(\omega) \equiv \tilde{G}(\omega)+\tilde{G_c}(\omega) 
\end{eqnarray*}

\section{Fluorescence Spectrum}
\label{secFRS}
Because the two-time correlation function of the atomic dipole is proportional to the first order coherence function $g^{(1)}(\tau)$ \cite{MQO} of the radiated photon field and the fluorescence spectrum can be obtained by taking the Fourier transform of the first order coherence function, one has:
\begin{eqnarray} 
\label{sw}
S(\omega) = \int_{-\infty}^\infty d\tau\, g^{(1)}(\tau)e^{i\omega\tau} \propto \langle\tilde{\sigma}_+(\omega)\tilde{\sigma}_-(-\omega)\rangle_R 
\end{eqnarray} 
In this way the fluorescence spectrum can be easily determined from Eqs. (\ref{ss1}-\ref{ss2}) after determining the noise correlation functions.
It should be noted that here we cannot directly apply the quantum regression theorem since it is invalid for non-Markovian process. We avoid this difficulty by linearize the Bloch equations with the Liouville operator expansion and by solving the linearized equations directly in the frequency domain.

As a check, we first use our formulation to calculate the results for the free space case, where the memory functions are delta functions, i.e. $\sum_k|g_k|^2 e^{\pm i\Delta_k t}= \Gamma\delta(t)$ and $\Gamma$ is the decay rate of the excited atom.
The correlation functions of the noise operators at zero temperature are also delta-function correlated (white noises). 
Therefore, the fluorescence spectrum at steady-state is given by: 
\begin{widetext}
\begin{eqnarray} 
\label{swhite} 
\langle\tilde{\sigma_+}(\omega-\Delta)\tilde{\sigma_-}(-\omega+\Delta)\rangle_R &=& \frac{\pi^2\Omega^2(\frac{\Gamma^2}{4}+\Delta^2)}{A^2}\delta(\omega+\Delta)\\\nonumber
&+& \frac{\pi\Gamma\Omega^4(\frac{\Omega^2}{2}+\Gamma^2+(\omega+\Delta)^2)}{2 A\{\Gamma^2[A-2(\omega+\Delta)^2]^2+(\omega+\Delta)^2[\Omega^2+\Delta^2+\frac{5}{4}\Gamma^2-(\omega+\Delta)^2]^2\}} 
\end{eqnarray}
\end{widetext} 
where $A \equiv \frac{\Omega^2}{2}+\Delta^2+\frac{\Gamma^2}{4}$. In the limit of strong on-resonance pumping ($\Omega\gg\Gamma$, $\Delta=0$), Eq. (\ref{swhite}) can be reduced to: 
\begin{eqnarray}
\label{eqMollow} 
&&\hspace{-0.4cm}\langle\tilde{\sigma_+}(\omega)\tilde{\sigma_-}(-\omega)\rangle_R = 2\pi\cdot[2\pi\frac{\Gamma^2}{4\Omega^2}\delta ({\omega})+\\\nonumber
&&\hspace{-0.4cm}\frac{\frac{3}{16}\Gamma}{(\omega+\Omega)^2+\frac{9}{16}\Gamma^2}\nonumber+\frac{\frac{1}{4}\Gamma}{\omega^2+\frac{1}{4}\Gamma^2}+\frac{\frac{3}{16}\Gamma}{(\omega-\Omega)^2+\frac{9}{16}\Gamma^2}] 
\end{eqnarray} 

So far we have proved that the resonance fluorescence spectrum exhibits the Mollow's triplet spectral shape \cite{BMollow69}: three Lorentzian profiles with peaks in the ratio $1:3:1$, and widths of $\frac{3}{2}\Gamma$, $\Gamma$, and $\frac{3}{2}\Gamma$.
There is a contribution from the elastic Rayleigh scattering in the center part (which is a delta function with zero detuning frequency) and a contribution from the inelastic Raman scattering (the three peaked profiles).
The line-width of each peak is proportional to the decay-rate of atom 
and the separation of adjacent peaks is proportional to the Rabi frequency.
This check provides a good support for our new formulation. 

Next we come back to calculate the case for a photonic bandgap material.
By using the density of states for three-dimensional photonic bandgap crystals, $ D(\omega) = \frac{1}{A^{3/2}}\sqrt{\omega-\omega_c}\Theta(\omega-\omega_c)$, we get the following noise correlation functions in the frequency domain at zero 
temperature:
\begin{widetext}
\begin{eqnarray} 
\label{noisepbg1} 
\langle \tilde{n}_-(\omega_1)\tilde{n}_+(-\omega_2)\rangle_R &=& \pi N(\omega_1)\Theta(\omega_1+\omega_a-\omega_c)\delta(\omega_1-\omega_2)\\
\label{noisepbg2}
\langle \tilde{n}_z(\omega_1)\tilde{n}_z(-\omega_2)\rangle_R &=& N(\omega_1)[2\pi\delta(\omega_1-\omega_2)+\langle\tilde{\sigma}_z(\omega_1-\omega_2)\rangle_R]\Theta(\omega_1+\omega_a-\omega_c)\\
\label{noisepbg3} 
\langle \tilde{n}_z(\omega_1)\tilde{n}_-(-\omega_2)\rangle_R &=& 0 \\ 
\label{noisepbg4} 
\langle \tilde{n}_-(\omega_1)\tilde{n}_z(-\omega_2)\rangle_R &=& N(\omega_1)\langle\tilde{\sigma}_-(\omega_1-\omega_2)\rangle_R\Theta(\omega_1+\omega_a-\omega_c)\\
\label{noisepbg5}
\langle \tilde{n}_z(\omega_1)\tilde{n}_+(-\omega_2)\rangle_R &=& N(\omega_1)\langle\tilde{\sigma}_+(\omega_1-\omega_2)\rangle_R\Theta(\omega_1+\omega_L+-\omega_c)\\ 
\label{noisepbg6} 
\langle \tilde{n}_+(\omega_1)\tilde{n}_z(-\omega_2)\rangle_R &=& 0 
\end{eqnarray} 
\end{widetext}
where $N(\omega) \equiv 4\beta^{3/2}\frac{\sqrt{\omega_a+\omega-\omega_c}}{\omega_a+\omega}$.

\begin{figure}[t]
\begin{center} 
\includegraphics[width=3.0in]{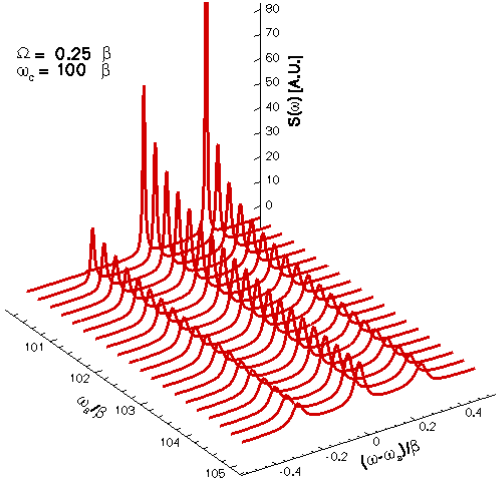} 
\caption{Resonance fluorescence spectra for $\omega_a$ outside the band edge $\omega_c$ at a constant Rabi frequency.}
\label{swpbg2}
\end{center} 
\end{figure} 

\begin{figure}[b]
\begin{center}
\includegraphics[width=3.0in]{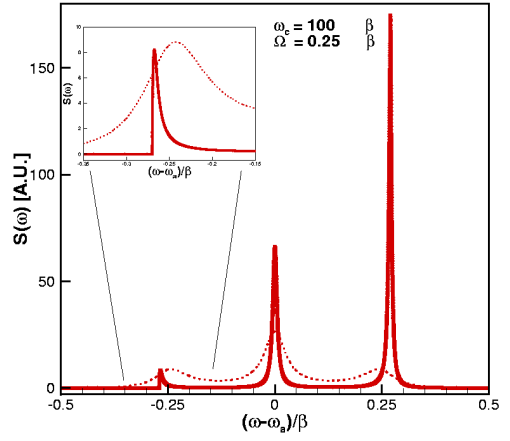}
\caption{Fluorescence spectra for $\omega_a$ near the band edge $\omega_c$. Solid line: $\omega_a = 100.27 \beta$. Dotted line: $\omega_a = 1000 \beta$. The inset is the enlarged profile of the lower frequency peak.}
\label{swpbg3}
\end{center} 
\end{figure} 

After applying these noise correlation functions in Eqs. (\ref{noisepbg1}-\ref{noisepbg6}) as well as the memory functions in Eqs. (\ref{Gpbg1}-\ref{Gpbg2}), we can get the fluorescence spectrum under resonant pumping from Eq. (\ref{sw}): 
\begin{widetext}
\begin{eqnarray}
\label{eqswpbg}
\langle\tilde{\sigma_+}(\omega)\tilde{\sigma_-}(-\omega)\rangle_R &=& \frac{4\pi^2\Omega^2 f_0 g_0 \tilde{G'}_0^2}{[\Omega^2(f_0+g_0)+2 f_0 g_0 h_0]^2}\delta(\omega)\\\nonumber
&+& N(\omega)\cdot\frac{\pi\Omega^4+i\Omega^3 g_{-\omega}\langle\tilde{\sigma}_-(0)\rangle_R-i\Omega^3 f_\omega\langle\tilde{\sigma}_+(0)\rangle_R+\Omega^2 f_\omega g_{-\omega}(2\pi+\langle\tilde{\sigma}_z(0)\rangle_R)}{[\Omega^2(f_\omega+g_\omega
)+2 f_\omega g_\omega h_\omega][\Omega^2(f_{-\omega}+g_{-\omega})+2 f_{-\omega} g_{-\omega} h_{-\omega}]}\Theta(\omega+\omega_a-\omega_c)
\end{eqnarray}
\end{widetext}

\begin{figure} 
\begin{center}
\includegraphics[width=3.0in]{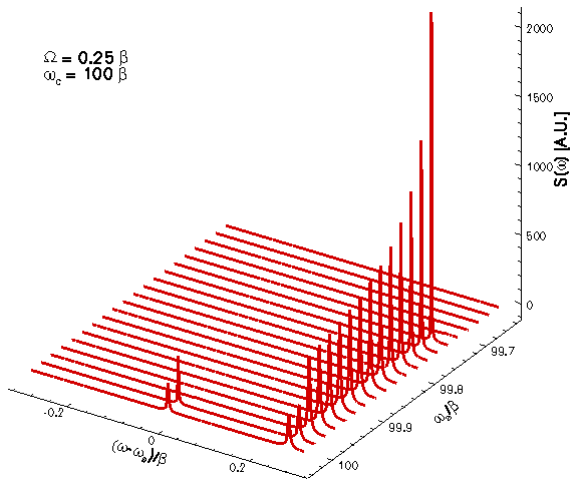}
\caption{Resonance fluorescence spectra for $\omega_a$ smaller than the bandedge $\omega_c$ at a constant Rabi frequency.}
\label{swpbg4} 
\end{center}
\end{figure}

By using Eq. (\ref{eqswpbg}), in Fig. \ref{swpbg2} we plot the resonance fluorescence spectrum at a constant Rabi frequency when the atomic transition frequency $\omega_a$ is above the bandedge $\omega_c$.
Here the linewidth of each peak is proportional to the normalized frequency $\beta$ defined bellow Eq.(\ref{Gpbg2}) and the separation of each adjacent peaks is also determined by the Rabi frequency.
The parameters used in the calculation are also labeled in the figure.
When the atomic transition frequency is far away from the band edge ($\omega_a\gg\omega_c$), the normal resonance fluorescence spectrum of Mollow's triplets is obtained just as expected.
As the atomic transition frequency moving toward the band edge, the profiles of incoherence scattering processes become sharper and sharper as there are fewer and fewer DOS available.
The profile in the lower frequency is not only suppressed but also asymmetrical due to the existence of the bandgap as shown in Fig. \ref{swpbg3}.
Its residual profile exhibits a sharp edge as shown in the inset of the figure.
It should also be noted that the peak in the higher frequency is enhanced a lot as can be clearly seen in the figure.
Eventually the peak in the lower frequency will be totally suppressed when the atomic transition frequency is moving more toward the band edge.
At this time the resonance fluorescence spectrum now only has two peaks, as shown in Fig. \ref{swpbg4}.
This is of course again due to the bandgap effect.
It is interesting to see that now the enhancement of the original middle frequency peak becomes smaller than the original higher frequency peak.
If the atomic transition frequency moves below the band edge, then it is no longer possible to resonantly pump the two level atom located within the photonic bandgap material.
Although in principle non-resonant pumping through the use of three-level atoms can be realized, the fluorescence spectra will be different when compared to the resonance fluorescence spectra considered in the present paper. This will be one of the interesting topics that can be further studied in the future along this line of research directions.        

\begin{figure}
\begin{center}
\includegraphics[width=3.0in]{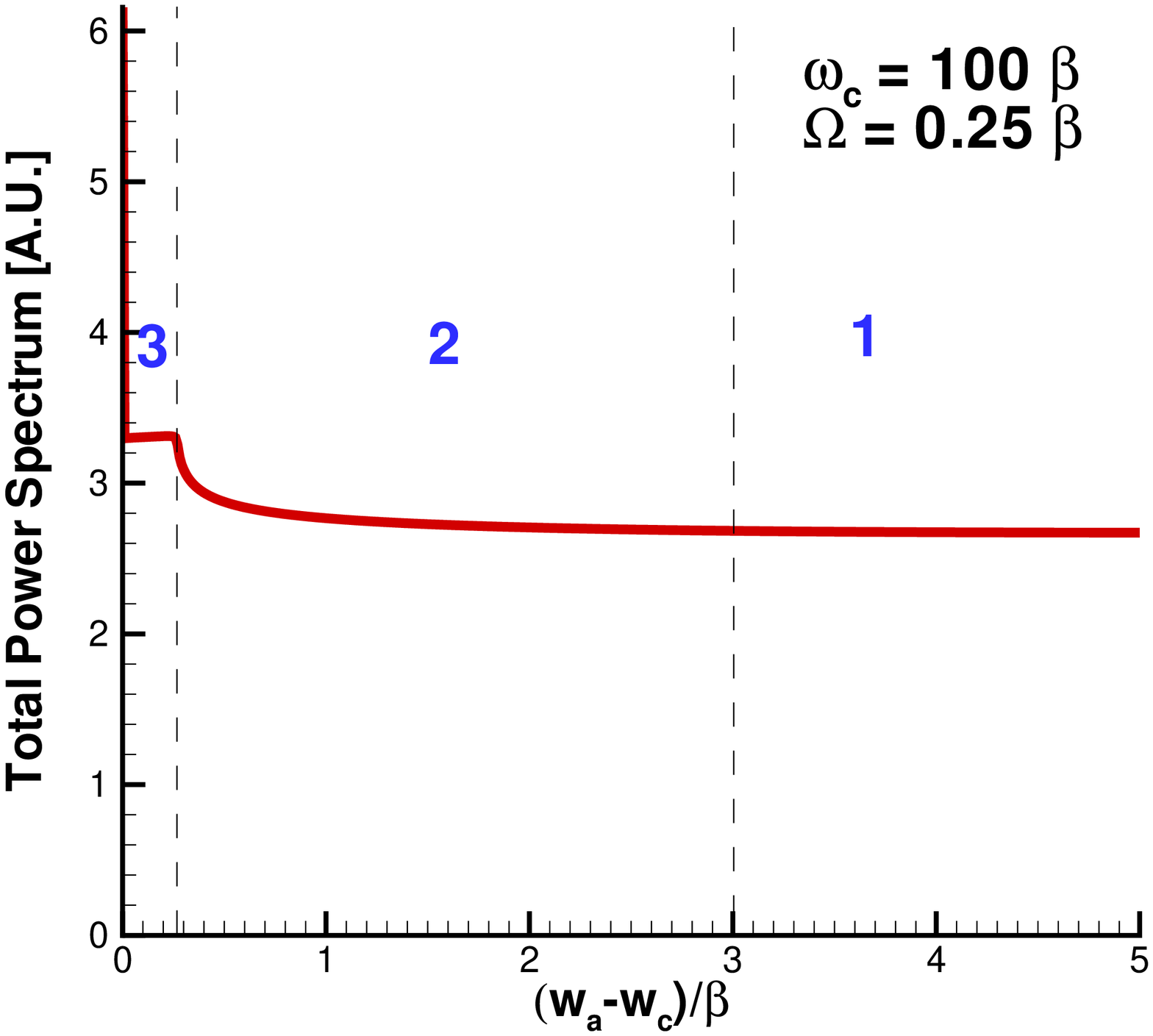}
\caption{Total fluorescence power spectrum for $\omega_a$ near the bandedge.} 
\label{figtsw}
\includegraphics[width=3.0in]{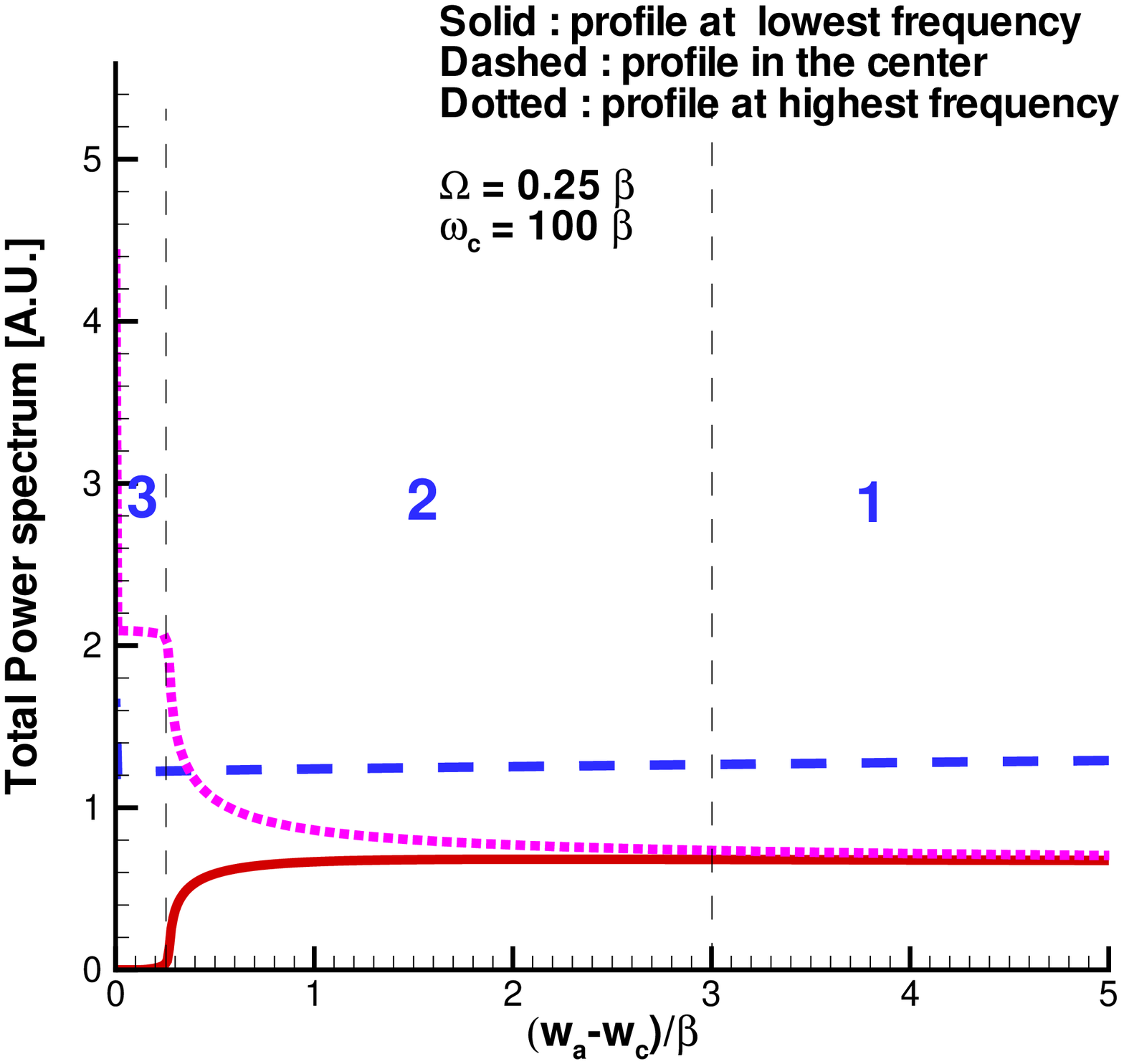}
\caption{Individual fluorescence powers for different peaks. Solid line: power of the lower frequency peak; Dashed line: power of the central frequency peak; Dotted line: power of the higher frequency peak.} 
\label{figtsw3}
\end{center}
\end{figure}

We also plot the total resonance fluorescence power spectrum in Fig. \ref{figtsw} by integrating over all frequencies.
The curve exhibits three regions of different behaviors for different offsets between the transition frequency and the bandedge frequency.  
When the transition frequency is far above the band edge, Region 1, the total power spectrum remains almost the same due to small variation of available DOS.
As one moves the transition frequency closer to the band edge, Region 2, the total power spectrum will increase as the photonic crystals begin to impact the inelastic scattering processes, such as the case in Fig. \ref{swpbg3}.
The total power spectrum will saturate when the first profile of Mollow's triplet is totally suppressed, Region 3.
For comparison, the optical power corresponding to each profile peak is also plotted in Fig. \ref{figtsw3} for different wavelength offsets.

\section{Conclusions} 
\label{secConclu}
In conclusion, we have developed a new formulation to calculate the fluorescence spectrum with non-Markovian photon-atom interactions, and have successfully applied it to the case of a single atom embedded in a photonic bandgap crystal.
By introducing Liouville operator expansion, we can overcome some of the difficulties caused by the non-Markovian nature of the problem due to the non-uniform distribution of the photon states in the photonic bandgap crystals. 
Although the approximation we have made is valid only when the time scale of the system response is much slower compared to the time scale of the memory function, however, with the use of our approximation we can still keep the finite response time of the memory function in the formulation.
This allows us to study the effects that are not considered in the usual Markovian regime, where the memory function is simply approximated by a delta-function.
Although our approximation includes only one portion of the non-Markovian nature of the problem, it should still be quite valid for the fluorescence spectrum calculation considered in the present work since here the memory function time scale is typically much shorter than the atomic response time scale.
We think one important contribution of the present work is to demonstrate that even in this unconsidered not-fully-non-Markovian regime there are still some interesting new phenomena like the suppression and enhancement of the relative fluorescence peak amplitudes at different wavelength offsets.
Historically, in 1975 the Mollow's triplets for the free space case was observed experimentally for the first time by using the hyperfine structure of the sodium atom \cite{FWu75}. 
It shall be very interesting to see if one can actually observe the predicted phenomena experimentally in the near future. We also believe that the results we have found should open up many new studies along this line of research directions.

\begin{acknowledgments}
One of the authors (R. Lee) wishes to acknowledge valuable conversations with Dr. Zheng Yao Su and Dr. Yu-Li Lee.
\end{acknowledgments}

\appendix*
\section{First Order Liouville Expansion}
\label{sec1st}
In this appendix we carry out the Liouville operator expansion in Eq. (\ref{Liouville}) to the first order:
\begin{eqnarray}
H \approx H_0 =\frac{\hbar}{2}(\sigma_- e^{i\Delta t}+\sigma_+ e^{-i\Delta t})+\hbar\sum_k\omega_k a_k^\dag a_k
\end{eqnarray}
The atom operators $\sigma_z(t)$, $\sigma_-(t)$, and $\sigma_+(t)$ can then be approximated by
\begin{widetext}
\begin{eqnarray}
\sigma_z(t) &\approx& \cos[\Omega(t-t')]\sigma_z(t')+i\sin[\Omega(t-t')][\sigma_-(t')e^{i\Delta t'}-\sigma_+(t') e^{-i\Delta t'}]\\
\sigma_-(t) &\approx& \frac{i e^{-i\Delta t'}}{2}\sin[\Omega(t-t')]\sigma_z(t')\\\nonumber
&+& \frac{e^{-i\Delta t'}}{2}\cos[\Omega(t-t')][\sigma_-(t')e^{i\Delta t'}-\sigma_+(t') e^{-i\Delta t'}]+\frac{1}{2}[\sigma_-(t')+e^{-2 i\Delta t'}\sigma_+(t')]\\
\sigma_+(t) &\approx& \frac{-i e^{i\Delta t'}}{2}\sin[\Omega(t-t')]\sigma_z(t')\\\nonumber
&-&\frac{e^{i\Delta t'}}{2}\cos[\Omega(t-t')][\sigma_-(t')e^{i\Delta t'}-\sigma_+(t') e^{-i\Delta t'}]+\frac{1}{2}[\sigma_+(t')+e^{2 i\Delta t'}\sigma_-(t')]
\end{eqnarray}
\end{widetext}
It can easily be seen that above equations are reduced to the zero-th order equations when the Rabi frequency $\Omega$ is small.
With the first order Liouville operator expansion, the generalized optical Bloch equations in Eqs. (\ref{es1}-\ref{es3}) become:
\begin{widetext}
\begin{eqnarray}
&&\hspace{-0.4cm}\dot{\sigma}_-(t) = i\frac{\Omega}{2}\sigma_z e^{-i\Delta t}-\int_{-\infty}^t d\,t' G(t-t')\cos[\Omega(t-t')]\sigma_-(t')- i\int_{-\infty}^t d\,t' G(t-t')\sin\Omega(t-t')](\frac{1+\sigma_z(t')}{2})\\
&&\hspace{-0.4cm}\dot{\sigma}_+(t) = -i\frac{\Omega}{2}\sigma_z e^{i\Delta t} -\int_{-\infty}^t d\,t' Gc(t-t')\cos[\Omega(t-t')]\sigma_+(t')+ i\int_{-\infty}^t d\,t' Gc(t-t')\sin\Omega(t-t')](\frac{1+\sigma_z(t')}{2})\\
&&\hspace{-0.4cm}\dot{\sigma}_z(t) = i\Omega (\sigma_-e^{i\Delta t} -\sigma_+ e^{-i\Delta t})- \int_{-\infty}^t d\,t' [G(t-t')+Gc(t-t')](\frac{1+\sigma_z(t')}{2})\\\nonumber
&&\hspace{0.4cm}- i \int_{-\infty}^t d\,t'\sin[\Omega(t-t')\{G(t-t')\sigma_-(t')e^{i\Delta t'}- Gc(t-t')\sigma_+(t')e^{-i\Delta t'}\}\\\nonumber
&&\hspace{0.4cm}-\int_{-\infty}^t d\,t'\cos[\Omega(t-t')][G(t-t')+Gc(t-t')](\frac{1+\sigma_z(t')}{2})
\end{eqnarray}
\end{widetext}

By the same method of Fourier transform, we can get the fluorescence spectrum with following non-zero correlation functions of the noise operators at zero temperature:
\begin{widetext}
\begin{eqnarray}
\label{noisepbg11}
\langle\tilde{n}_-(\omega_1)\tilde{n}_+(-\omega_2)\rangle_R &=& N_1(\omega_1)[2\pi\delta(\omega_1-\omega_2)+\langle\tilde{\sigma'}_{-(\omega_1-\omega_2)}\rangle_R+\langle\tilde{\sigma'}_{+(\omega_1-\omega_2)}\rangle_R]\Theta(\omega_1+\omega_a+\Omega-\omega_c) \\\nonumber
&+& N_2(\omega_1)[2\pi\delta(\omega_1-\omega_2)-\langle\tilde{\sigma'}_{-(\omega_1-\omega_2)}\rangle_R-\langle\tilde{\sigma'}_{+(\omega_1-\omega_2)}\rangle_R]\Theta(\omega_1+\omega_a-\Omega-\omega_c)\\
\label{noisepbg12}
\langle\tilde{n}_z(\omega_1)\tilde{n}_z(-\omega_2)\rangle_R &=& N_1(\omega_1)[2\pi\delta(\omega_1-\omega_2)+2\langle\tilde{\sigma'}_{-(\omega_1-\omega_2)}\rangle_R+\langle\tilde{\sigma}_z(\omega_1-\omega_2)\rangle_R]\Theta(\omega_1+\omega_a+\Omega-\omega_c)\\\nonumber
&+& N_2(\omega_1)[2\pi\delta(\omega_1-\omega_2)-2\langle\tilde{\sigma'}_{-(\omega_1-\omega_2)}\rangle_R+\langle\tilde{\sigma}_z(\omega_1-\omega_2)\rangle_R]\Theta(\omega_1+\omega_a-\Omega-\omega_c)\\\nonumber
&+&\beta^{3/2}\frac{\sqrt{\omega_a+\omega_1-\omega_c}}{\omega_a+\omega_1}[4\pi\delta(\omega_1-\omega_2)+2\langle\tilde{\sigma}_z(\omega_1-\omega_2)\rangle_R]\Theta(\omega_1+\omega_a-\omega_c)\\
\label{noisepbg14}
\langle\tilde{n}_-(\omega_1)\tilde{n}_z(-\omega_2)\rangle_R &=& N_1(\omega_1)[2\pi\delta(\omega_1-\omega_2-\Delta)+2\langle\tilde{\sigma}_-(\omega_1-\omega_2)\rangle_R+\langle\tilde{\sigma}^-_{z(\omega_1-\omega_2)}\rangle_R]\Theta(\omega_1+\omega_a+\Omega-\omega_c)\\\nonumber
&+& N_2(\omega_1)[-2\pi\delta(\omega_1-\omega_2+\Delta)+2\langle\tilde{\sigma}_-(\omega_1-\omega_2)\rangle_R-\langle\tilde{\sigma}^-_{z(\omega_1-\omega_2)}\rangle_R]\Theta(\omega_1+\omega_a-\Omega-\omega_c)\\
\label{noisepbg15}
\langle\tilde{n}_z(\omega_1)\tilde{n}_+(-\omega_2)\rangle_R &=& N_1(\omega_1)[2\pi\delta(\omega_1-\omega_2+\Delta)+2\langle\tilde{\sigma}_+(\omega_1-\omega_2)\rangle_R+\langle\tilde{\sigma}^-_{z(\omega_1-\omega_2)}\rangle_R]\Theta(\omega_2+\omega_a+\Omega-\omega_c)\\\nonumber
&+& N_2(\omega_1)[-2\pi\delta(\omega_1-\omega_2+\Delta)+2\langle\tilde{\sigma}_+(\omega_1-\omega_2)\rangle_R-\langle\tilde{\sigma}^+_{z(\omega_1-\omega_2)}\rangle_R]\Theta(\omega_2+\omega_a-\Omega-\omega_c)
\end{eqnarray}
\end{widetext}
where 
\begin{eqnarray*}
N_1(\omega) &\equiv& \beta^{3/2}\frac{\sqrt{\omega_a+\omega+\Omega-\omega_c}}{\omega_a+\omega+\Omega}\\
N_2(\omega) &\equiv& \beta^{3/2}\frac{\sqrt{\omega_a+\omega_1-\Omega-\omega_c}}{\omega_a+\omega_1-\Omega}\\
\tilde{\sigma'}_{\pm(\omega_1-\omega_2)} &\equiv& \tilde{\sigma}_\pm(\omega_1-\omega_2\mp\Delta)\\
\tilde{\sigma}^{\pm}_{z(\omega_1-\omega_2)} &\equiv& \tilde{\sigma}_z(\omega_1-\omega_2\pm\Delta)
\end{eqnarray*}
>From Eq. (\ref{noisepbg11}-\ref{noisepbg15}), the correction of high order terms will have impacts on the fluorescence spectrum only when the Rabi frequency $\Omega$ is of the same magnitude with other characteristic frequencies of the system, i.e. $\omega_a$, and $\omega_c$.
For optical systems we are interested in, the zero-th order Liouville operator expansion shall give us reasonable results.
Only for the cases of very strong pumping power or ultra short atom lifetime we need to take into account the higher order corrections.

\end{document}